\documentstyle[aps,prl,preprint]{revtex}
\begin{document}

\draft

\title{Floquet Analysis of Atom Optics Tunneling Experiments}
\author{Robert Luter and L.~E. Reichl}
\address{Center for Studies in Statistical Mechanics and Complex
Systems, The University of Texas at Austin,
Austin, Texas 78712}
\date{August 27, 2002}
\maketitle

\begin{abstract}

Dynamical tunneling has been observed in atom optics experiments by two
groups. We show that the
experimental
results are extremely well described by time-periodic Hamiltonians with momentum quantized
in units of the atomic recoil. The
observed tunneling has a well defined period when only two Floquet states
dominate the dynamics.  Beat frequencies are observed when three Floquet states dominate.
We find  frequencies which match those observed in both experiments. The dynamical
origin of the dominant Floquet states
is identified.
\end{abstract}


\bigskip
\bigskip

Atom optics experiments recently have been used to investigate the effect of
underlying classical chaos on quantum dynamics. The experiments we focus on
in this paper have demonstated the existence of dynamic tunneling in momentum
space in regimes where the underlying classical phase space contains a
mixture of chaotic and regular orbits. We will show that we can accurately
reproduce the dominant tunneling frequencies observed in these two very
different experiments using Floquet analysis of the quantum dynamics.

Typically, cold
sodium or cesium atoms are allowed to interact with laser beams which are detuned away
from resonance with two atomic energy levels which have energy spacing,
${\hbar}{\omega}_0$. Two counterpropagating laser beams create a periodically
modulated standing wave of light which stimulates absorption and then emission of a
photon.
This results in a net atomic recoil of $2\hbar k_{L}$, where $k_{L}=\omega
_{L}/c$ is
the wave vector of the laser beams and $\hbar$ is Planck's constant. When
the laser
detuning, $\delta _{L}=\omega _{0}-\omega _{L}$ is large, this process
dominates the
dynamics.

A theoretical model which describes the atomic dynamics in such systems was
developed by Graham, Schlautmann, and Zoller \cite{Graham}.
  Recently, two groups,
Steck, Oskay and Raizen \cite{Steck1,DSteck2} in Texas and Hensinger et.al.
at NIST
\cite{Hensinger},
have performed independent experiments
in which dynamic tunneling has been observed. In this
letter,  we explore the accuracy of the models used to analyse these
experiments, and the dynamical origin of the tunneling  observed in each
experiment. We first discuss the Texas experiment and then the NIST
experiment.

In the Texas experiment \cite{Steck1,DSteck2}, the dynamics of
non-interacting cold
cesium atoms, in an amplitude modulated standing wave of light, was measured.
The atomic center-of-mass Hamiltonian (in S.I. units) used to model dynamics of the
cesium atoms is
\begin{equation}
\widehat{H}=\frac{\widehat{p}^{2}}{2m}-2V_{o}\cos ^{2}\left(
\frac{\omega _{m}t}{2}\right) \cos \left( 2k_{L}\widehat{x}\right),
\label{texasSIham}
\end{equation}
where ${\hat p}$,  ${\hat x}$, and $m$ are the momentum, position, and mass,
respectively, of a cesium atom,
$\omega _{m}=\frac{2\pi }{T}$ is the modulation frequency, and
$V_{o}=\frac{\hbar \Omega ^{2}_{max}}{8\delta _{L}}$ is the ac Stark shift
amplitude, where $\Omega_{max}=-2E_0d/{\hbar}$ is the Rabi frequency, $E_0$ is
the electric field strength, and $d$ is the dipole moment of cesium
\cite{Steck3}.

In the experiments,  the initial state is well localized at discrete momentum states
separated by $2{\hbar}k_L$.   This
quantization of the momentum occurs naturally in the experiment due to the presence of
counter propagating laser beams which cause two-photon transitions. Therefore, in our
theoretical analysis, we perform a  scaling which explicitly  quantizes the momentum
in units of $2{\hbar}k_L$. As we will see later, this allows us to use Floquet theory
rather than  Floquet-Bloch theory deals with a continuum of monentum states 
\cite{Mouchet}. Let
${\widehat{\phi}}=2k_L{\widehat{x}}$,
${\widehat{p}}=2{\widehat{n}}{\hbar}k_L$,
${\omega}_r={\hbar}k_L^2/2m$, ${\omega}={\omega}_m/4{\omega}_r$,
$t'=4{\omega}_rt$, and ${\widehat{H}}_{th}=m{\widehat{H}}/2k_L^2{\hbar}^2$, to
obtain,
\begin{equation}
\widehat{H}_{th}
= \widehat{n}^2-{{\alpha}{\omega}^2\over 8{\pi}^2}{\biggl[}
{\cos}(\widehat{\phi}) +{1\over 2}{\cos}(\widehat{\phi}-{\omega}t') +{1\over
2}{\cos}(\widehat{\phi}+{\omega}t') {\biggr]},
\label{texasTHham}
\end{equation}
where ${\alpha}=8{\omega}_rT^2V_0/{\hbar}$. All quantities are dimensionless and ${\hat
n}$ is the dimensionless momentum
operator with eigenstates,
$|n{\rangle}$, and integer eigenvalues, $-{\infty}{\leq}n{\leq}\infty$.
(Note that the experimental papers \cite{Steck1,DSteck2} perform the 
following scaling,
${\widehat{\phi}}=2k_L{\widehat{x}}$,
${\tau}={\omega}_mt/2{\pi}=t/T$,
${\widehat{\rho}}=4{\pi}k_L{\widehat{p}}/m{\omega}_m$,
${\widehat{H}}_{exp}=16{\pi}^2k_L^2{\widehat{H}}/m{\omega}_m^2$, and
Hamiltonian  takes the form,
$\widehat{H}_{exp}={\widehat{\rho}^2\over
2}-2{\alpha}{\cos}^{2}({\pi}{\tau}){\cos}(\widehat{\phi})$.)  This system has three
primary resonances centered at
$(n=0,
~{\phi}=0)$  and  $(n={\pm}{\omega}/2,~{\phi}=0)$. For small
values of  $\alpha$ (${\alpha}<1.5$), the primary resonances  have
pendulum-like structure, and the resonance at $n=0$ has half-width
${\Delta}n_0=\sqrt{{\alpha}{\omega}^2\over
4{\pi}^2}$, while the resonances at  $n_{\pm}={\pm}{\omega}/2$ have half-width
${\Delta}n_{\pm}={\Delta}n_0/\sqrt{2}$ \cite{Reichl}. The primary resonance at $n=0$
bifurcates at
${\alpha}{\approx}7.0$. The two outer primaries remain visible until
${\alpha}{\approx}13.0$ when they disappear.

The classical motion is obtained from Hamilton's equations,
$\dot{n} = -\frac{\partial H_{th}}{\partial \phi }$ and $\dot{\phi }
=\frac{\partial H_{th}}{\partial n}$.  In the Texas
experiment,
${\omega}_{r}=1.30{\times}10^{4}rad/s$ and $T=2{\pi}/{\omega}_m=20{\mu}s$, which, for
small
$\alpha$, gives a location of
$n_{\pm} =\pm 3.0$ for the outer primary resonances. For the field strength
${\alpha}=9.7$, used in the Texas experiment, the
pendulum approximation for the half-widths of the two outer primary
resonances gives
${\Delta}n_{\pm}=2.1$, while the half-width
of the central island is ${\Delta}n_{0}=3.0$. Thus the Texas experiment, for
${\alpha}=9.7$, is in the strong field regime, where the primary resonances
have  overlapped and considerable chaos is expected \cite{Reichl}. A surface of
section of the classical phase space for
${\alpha}=9.7$ is shown in Figure (1.a). The central primary resonance has
bifurcated
and is
largely destroyed, and the outer primary resonances have been reduced
significantly in
size and
are centered at  momentum values  $n={\pm}4.2$. Note also that the chaotic
region lies in the interval $-5{\leq}n{\leq}+5$, indicating that eleven
quantized
momentum states determine the dynamics in the chaotic region.

The Texas experiment used atoms prepared initially with a narrow momentum
distribution peaked at
$n=4.2$ (on the upper island). To numerically simulate
this initial condition, we solved the Schrodinger equation,
$i{{\partial}|{\Psi}(t'){\rangle}\over {\partial}t'}={\hat
H}_{th}|{\Psi}(t'){\rangle}$, using momentum states,
$|n{\rangle}$, as a basis. A coherent state,
\begin{equation}
{\langle}n|{\Psi}(0){\rangle}{\equiv}\left\langle n\left| \phi
_{o}n_{o}\right.
\right\rangle =\left(
\frac{\sigma ^{2}}{\pi }\right) ^{\frac{1}{4}}\exp \left[
\frac{-\sigma ^{2}}{2}\left( n-n_{o}\right) ^{2}-i\left(
n-n_{o}\right) \phi _{o}\right]
\label{tex:initial}
\end{equation}
centered at ($n=n_o,{\phi}={\phi}_o$) is used as the initial state, with
${\sigma}=1.2$ which was used in the experiment. In the momentum basis, the Schrodinger
equation  reduces to a system
of coupled first order differential equations for the amplitudes,
${\langle}n|{\Psi}(t){\rangle}$.  This system was truncated, and 81
equations for the states ${\langle}n|{\Psi}(t){\rangle}$ with
$-40{\leq}n{\leq}40$ were kept. The time variation of the average momentum,
${\langle}n{\rangle}$, is shown in Figures (2.a), (2.b), and (2.c) for
\( \alpha = 8.0, 9.7, \) and \( 13.0 \), respectively. In all cases, the
initial state is ($n_o=4.2,{\phi}_o=0$). For all three plots, the average momentum
oscillates between the outer primary resonances.
The plot for  $\alpha=8.0$ ($\alpha=9.7$) has two dominant frequencies,
$f_1=1.95kHz$ and $f_2=2.73kHz$ ($f_1=2.39kHz$ and $f_2=2.88kHz$),
giving rise
to a beating effect. The beating effect at $\alpha=9.7$ was observed in the
Texas
experiment \cite{DSteck2}, but the experimental error bars were too great to
resolve
it at $\alpha=8.0$. The plot for $\alpha=13.0$ shows one dominant frequency,
$f=1.56kHz$.

It is useful to examine these results using Floquet theory \cite{Reichl}.
Because the Hamiltonian, ${\hat H}_{th}$, has time periodic coefficients, the
Schrodinger equation has Floquet solutions of the form
${\langle}n|{\Psi}(t){\rangle}={\rm
e}^{-i{\Omega}_jt}{\langle}n|{\chi}_j(t){\rangle}$ where ${\Omega}_j$ is the
$j^{th}$ Floquet eigenphase and $|{\chi}_j(t){\rangle}$ is the $j^{th}$
Floquet eigenstate and is periodic in time,
$|{\chi}_j(t){\rangle}=|{\chi}_j(t+T){\rangle}$ \cite{Reichl}. The Floquet
eigenphases,
${\Omega}_j$,  are conserved quantities, and the eigenstates form a complete
orthonormal basis which can be used to analyze the dynamics. The states,
$|{\chi}_j(0){\rangle}$, are eigenfunctions of the Floquet matrix, ${\hat
U}(T)$, and the phase functions, ${\rm e}^{-i{\Omega}_jT}$, are its
eigenvalues.
The Floquet matrix is computed by
by taking a momentum eigenstate as the initial state and evolving it for
one period,
$T$, using the Schrodinger equation. The resulting vector (in the momentum
basis) is a
column of the Floquet matrix.

The overlap probabilities,
$P_j{\equiv}|{\langle}{\chi}_j(0)|{\phi}_on_o(0){\rangle}|^2$,   give the
contribution of each Floquet state to the dynamics.
The probability to find the system in momentum state, $|n{\rangle}$, at time
$t$, can be written \cite{Reichl}
\begin{eqnarray}
\left| \left\langle n\left| \phi
_{o}n_{o}\left( t\right) \right. \right\rangle \right| ^{2}
& = & \sum _{i}\sum _{j}\exp \left( -i\left( \Omega _{j}-\Omega _{i}\right)
t\right)
\left\langle n\right. \left| \chi_{j}\left( t\right) \right\rangle
\left\langle \chi_{i}\left( t\right) \right. \left| n\right\rangle
\nonumber \\
& & \times \left\langle \chi _{j }\left( 0\right) \left| \phi
_{o}n_{o}\left( 0\right) \right. \right\rangle \left\langle \phi
_{o}n_{o}\left(
0\right) \left| \chi _{i}\left( 0\right) \right. \right\rangle,
\label{eq:9}
\end{eqnarray}
with time $t$ in seconds and ${\Omega}_j/2\pi$ in Hertz.
The oscillation frequencies, $f_{exp}$, observed in the experiments can
be equated to differences between  Floquet eigenphases.
The frequency differences, $f_{exp}=(\Omega
_{j}-\Omega _{i})/2{\pi}$, for Floquet eigenstates with
overlap probability, $P_iP_j{\geq}0.04$,  are plotted in Figure (3) for the
range of parameters shown in the Texas experiment \cite{DSteck2}. Each curve is a plot
of the frequency difference between two Floquet states as a function of $\alpha$. At
values of $\alpha$ where there are multiple curves there are more than two dominant
frequencies.  The Texas experiment was able to resolve the dominant frequencies,
$f_{exp}<3$kHz,  in the
interval between 
${\alpha}{\approx}8.7$ and ${\alpha}{\approx}10.3$. Our analysis exactly
reproduces those experimental results.  In the amplitude range,
${\alpha}{\approx}7.6$ to ${\alpha}{\approx}11.6$, we find that
two frequencies dominate and give rise to the beats seen in Figures
(2.a) and (2.b).   In the Texas data \cite{DSteck2}, large error
bars occur
in the regions ${\alpha}={\leq}7.0$ and ${\alpha}{\geq}13.7$. This may be
due to the
rapid change in the dominant frequencies in those regions.  A
fundamental change
in the dynamics occurs for ${\alpha}>14$, where a different set of Floquet
states begins to dominate the dynamics.

Only eleven Floquet states have support on momentum in the region, $n=-5$ to $n=5$,
and determine the dynamics in the chaotic region. In Figures (4.a)-(4.d),  we show
Husimi plots for the Floquet states which, for
${\alpha}=9.7$,  have the largest
overlap probability, three of which dominate the dynamics.  The dark regions of these
plots show the region of the classical phase space where the probability of finding
the cesium atoms is largest.  The eigenphase differences,
$({\Omega}_{4b}-{\Omega}_{4a})/2\pi=2.89k$Hz and
$({\Omega}_{4a}-{\Omega}_{4c})/2\pi=2.40k$Hz correspond to the two dominant
oscillation frequencies observed by the Texas experiment at ${\alpha}=9.7$. The
state in Figure (4.d) has the fourth highest overlap probability, $P_d=0.045$,
but it lies in the chaotic sea. The state in Fig. (4d) and others not shown contribute
to the fine scale structure in these curves.

Let us now consider the NIST experiment \cite{Hensinger}, which used a
Bose-Einstein condensate of sodium atoms to observe dynamic tunneling. Formation of
a condensate with the sodium atoms yields a narrower distribution of initial
momenta than the Texas experiment. The Hamiltonian used to describe the experiment can
be written in the form 
\begin{equation}
\widehat{H}_{th}
=  \widehat{n}^2+{{\tilde {\omega}}^2{\kappa}\over 2}{\biggl[} {1+2\nu \epsilon
\cos}({\tilde {\omega}}
t') -\cos\widehat{\phi} - \nu {\epsilon}{\cos}(\widehat{\phi}-{\tilde {\omega}}t')
- \nu {\epsilon}{\cos}(\widehat{\phi}+{\tilde {\omega}}t') {\biggr]},
\label{nistTHham}
\end{equation}
where ${\nu} = {\pm} 1$. When $\nu=+1$ ($\nu=-1$), Eq. (\ref{nistTHham}), with
starting time $t'=0$, reproduces the dynamics of the NIST experiment which has
starting time, ${\tilde {\tau}}= {\tilde T}/4$ (${\tilde {\tau}}=3{\tilde T}/4$). 
This Hamiltonian again discretizes the momentum in units of $2\hbar {\tilde k}_L$,
which reflects the quantization of momentum due to the two photon transitions. 
(The Hamiltonian,  $\widehat{H}_{exp}
=\frac{\widehat{\rho}^{2}}{2}+ 2
\kappa
\left(1 + 2
\epsilon \sin({\tilde {\omega}}_{m}{\tilde {\tau}})\right) \sin ^{2}
{\bigl(}\frac{\widehat{\phi}}{2}{\bigr)}$,  used in the experimental paper is obtained
by setting 
$\widehat{\rho}={4{\hbar}{\tilde k}_L^2\over m{\tilde {\omega}}_m}{\hat n}$ and 
${\widehat{H}}_{exp}={8{\tilde k}_L^4{\hbar}^2\over
m^2{\tilde {\omega}}_m^2}{\widehat{H}}_{th}$.) 

  For small amplitudes, $\kappa$
and ${\kappa}{\epsilon}$, the NIST Hamiltonians have three primary
resonances. For
${\nu}=-1$ they are located at $(n=0, ~{\phi}=0)$
and
$(n={\pm}{{\tilde {\omega}}\over 2},~{\phi}={\pm}\pi)$, while for ${\nu}=+1$ they
are located
at
$(n=0, ~{\phi}=0)$ and
$(n={\pm}{{\tilde {\omega}}\over 2},~{\phi}=0)$. They have half-widths,
${\Delta}n_0=\sqrt{{\tilde {\omega}}^2{\kappa}}$ and
${\Delta}n_{\pm}=\sqrt{{\tilde {\omega}}^2{\kappa}{\epsilon}}$ \cite{Reichl}.

A strobe plot of the classical phase space for the Hamiltonian in Eq.
(\ref{nistTHham}) with ${\nu}=-1$ and experimental parameters
${\tilde {\omega}}_m/2\pi=250$kHz, ${\tilde {\omega}}=2.5$, ${\kappa}=1.66$ and
${\epsilon}=0.29$ is shown in Figure (1.b). Seven
Floquet states determine the dynamics in the chaotic region between
$n=-3$ and
$n=3$. For the parameters used in the experiment, the pendulum approximation
predicts the primary resonances
to lie at $n=0$ and $n={\pm}1.25$, and  have half-widths, ${\Delta}n_0=3.2$
and
${\Delta}n_{\pm}=1.7$.  We find that the primary resonances are
totally destroyed at ${\kappa}{\approx}0.2$, and then new resonances, which
resemble
the primaries, reappear and disappear repeatedly as $\kappa$ is increased. For
${\kappa}=1.66$ and
${\epsilon}=0.29$, a large resonance exists at $(n=0, \phi=0)$ and three
  small pairs of 
higher order resonances exist at
$(n{\approx}{\pm}1.5, \phi={\pm}\pi)$,  $(n{\approx}{\pm} 3.0,~{\phi}=0)$ and
$(n{\approx}{\pm} 2.0,~{\phi}=0)$.

In Figure (5) we show the time evolution of the momentum expectation value
for two
different initial conditions for the ${\nu}=-1$ Hamiltonian at parameter
values,
$\kappa=1.66$, $\epsilon=0.29$, ${\tilde {\omega}}=2.5$
and ${\tilde {\omega}}_m/2\pi=250$kHz.  Figure (5.a), with
$(n_o = 1.6, ~{\phi}_o=0)$, shows a somewhat noisy oscillation
with a dominant frequency  $24.9kHz$ (10.0 modulation periods), which is in good
agreement with
the experimental result.  Figure (5.b) shows the case with
  $(n_o = 3.0,~{\phi}_o=0)$. A clean oscillation with frequency
$18.3kHz$ (13.7 modulation periods) occurs. This oscillation was not observed in the
experiment, but we expect it would show up in a power spectrum of the experimental
data.

We now consider a Floquet analysis for both
Hamiltonians, ${\nu}={\pm}1$. The Floquet eigenphases for ${\nu}={\pm}1$ are
identical, but the Floquet eigenstates associated with each eigenphase
are different for the two Hamiltonians. Let us first consider the
$\nu=-1$ Hamiltonian with parameters, $\kappa=1.66$, $\epsilon=0.29$,
${\tilde {\omega}}=2.5$
and ${\tilde {\omega}}_m/2\pi=250$kHz. In
Figures (6.a) and (6.b), we show  the two Floquet states which dominate the
dynamics for  initial condition, $(n_o=1.6,~{\phi}_o=0)$.
They have a frequency difference, $({\Omega}_{6b}-{\Omega}_{6a})/2\pi=25.0$kHz.
Their frequency difference accounts for the
oscillation of 10 modulation periods  reported in
\cite{Hensinger}.  These Floquet states are not even-odd pairs as suggested in
\cite{Hensinger},  and they both lie in the chaotic sea.  If the
effective Planck's constant for this
experiment were smaller, more Floquet states would be supported by the
chaotic region and we would not expect to find this simple oscillation
\cite{timberlake} for this initial condition.

If we take initial condition, $(n_o=3.0,~{\phi}_o=0)$  for ${\nu}=-1$, we
obtain the oscillation shown in Figure (5.b). This oscillation  results
from the even-odd Floquet pair  shown in Figures (6.c) and (6.d). Figure (6.c) (Fig.
(6.d)) is even (odd) under the transformation $p{\rightarrow}-p$.  They have a
frequency difference,
$({\Omega}_{6c}-{\Omega}_{6d})/2\pi=18.3$kHz. This oscillation
appears to result from states sitting the outer-most nonlinear
resonance.

We finally consider the ${\nu}=+1$ Hamiltonian with parameters, $\kappa=1.66$,
$\epsilon=0.29$, ${\tilde {\omega}}=2.5$ and ${\tilde {\omega}}_m/2\pi=250$kHz. We
find that the
$25.0kHz$ (10 modulation periods) oscillation dominates those initial momentum
states which are centered at $\phi=0$ and lie in the interval $n_o=1.7$ to
$n_o=3.0$. These oscillations appear to result from the two Floquet states
which lie in the chaotic sea. If we change the parameters to
${\kappa} = 1.82$ and
${\epsilon} = 0.30$ and the modulation frequency to ${\tilde {\omega}}_m =
222kHz$, the dominant frequency for initial state,
$(n_o = 2.0,~{\phi}_o=0)$, is $36.8kHz$ (6.03 modulation periods), which is in
agreement with the NIST experiment.

In conclusion, the model Hamiltonians, with momentum quantized in units of
$2{\hbar}k_L$, give extremely good predictions of the experimental results.
Because of  the momentum quantization imposed by the dynamics of the experiment, we
found that it was advantageous to use Floquet theory rather than Floquet-Bloch theory
to analyse the experiment. In fact, our results are so good that these models might be
used to help calibrate future experiments.

\acknowledgments

The authors wish to thank the Welch Foundation, Grant
No.F-1051, NSF Grant INT-9602971  and DOE contract No.DE-FG03-94ER14405
for partial support of this work. We also thank the University of
Texas at Austin High Performance Computing Center for use of their computer
facilities, and we thank Mark Raizen, Dan Steck, Windell Oskay, and Chris
Helmerson for useful conversations.

\appendix

\begin{figure}
\caption{Classical strobe plots: (a) The Texas experiment with
${\omega}=6.0$ and \(\alpha = 9.7\). (b) NIST experiment with ${\omega}=2.5$,
\(\alpha = 1.66\) and ${\epsilon}=0.29$.}
\label{strobe3}
\end{figure}

\begin{figure}
\caption{Evolution of average momentum, 
${\langle}n{\rangle}$
(in dimensionless units) for the Texas experiment for ${\omega}=6.0$: (a)
${\alpha}=8.0$;  (b) ${\alpha}=9.7$;  and (c) ${\alpha}=13.0$.  }
\label{tex.mom}
\end{figure}

\begin{figure}
\caption{Oscillation frequencies,
${\Delta}{\Omega}=({\Omega}_j-{\Omega}_i)$, calculated
from the Floquet eigenphase differences for varying dimensionless field
strengths,
$\alpha$. A
threshold of $P_iP_j{\geq}0.04$ overlap probability was used to select the
dominant
frequencies. The three values shown at $\alpha=9.7$ correspond to 
$({\Omega}_{4a}-{\Omega}_{4b})/2{\pi}$, $({\Omega}_{4a}-{\Omega}_{4c})/2{\pi}$, and
$({\Omega}_{4b}-{\Omega}_{4c})/2{\pi}$}
\label{tex.freq}
\end{figure}

\begin{figure}
\caption{Husimi plots of Floquet eigenstates for the Texas
experiment for ${\omega}=6.0$ and
\(\alpha = 9.7\).  (a)  Floquet eigenphase
${\Omega}_{4a}/2\pi=16.9$kHz and
an overlap probability, $P_{4a}=0.416$. State (b) Floquet eigenphase
${\Omega}_{4b}/2\pi=19.7$kHz and an overlap probability, $P_{4b}=0.224$. (c)
Floquet eigenphase
${\Omega}_{4c}/2\pi=14.5$kHz and an overlap probability, $P_{4c}=0.20$. (d)
Floquet eigenphase ${\Omega}_{4d}/2\pi=18,389$Hz and an overlap probability,
$P_{4d}=0.045$.}
\label{tex.husimi}
\end{figure}

\begin{figure}
 \caption{Evolution of momentum expectation value,
${\langle}n{\rangle}$
(in dimensionless units) for the NIST experiment for $\kappa=1.66$,
$\epsilon=0.29$, ${\tilde {\omega}}=2.5$
and ${\tilde {\omega}}_m/2\pi=250$Hz: (a) $n_o=1.6$, and ${\phi}_o=0$;  (b)
$n_o=3.0$, and
${\phi}_o=0$.  }
\label{nist.mom}
\end{figure}

\begin{figure}
\caption{Husimi plots of Floquet eigenstates for the NIST
experiment
with $\kappa=1.66$, $\epsilon=0.29$, ${\tilde {\omega}}=2.5$
and ${\tilde {\omega}}_m/2\pi=250$Hz.  (a)  Floquet eigenphase
${\Omega}_{6a}/2\pi=49.0$kHz and overlap probability $P_{6a}=0.380$. State (b)
Floquet eigenphase
${\Omega}_{6b}/2\pi=73.9$kHz and overlap probability $P_{6b}=0.306$. (c) Floquet
eigenphase
${\Omega}_{6c}/2\pi=15.3$kHz and overlap probability $P_{6c}=0.427$. (d)
Floquet eigenphase ${\Omega}_{6d}/2\pi=33.5$kHz and overlap probability
$P_{6d}=0.421$.}
\label{nist.husimi}
\end{figure}

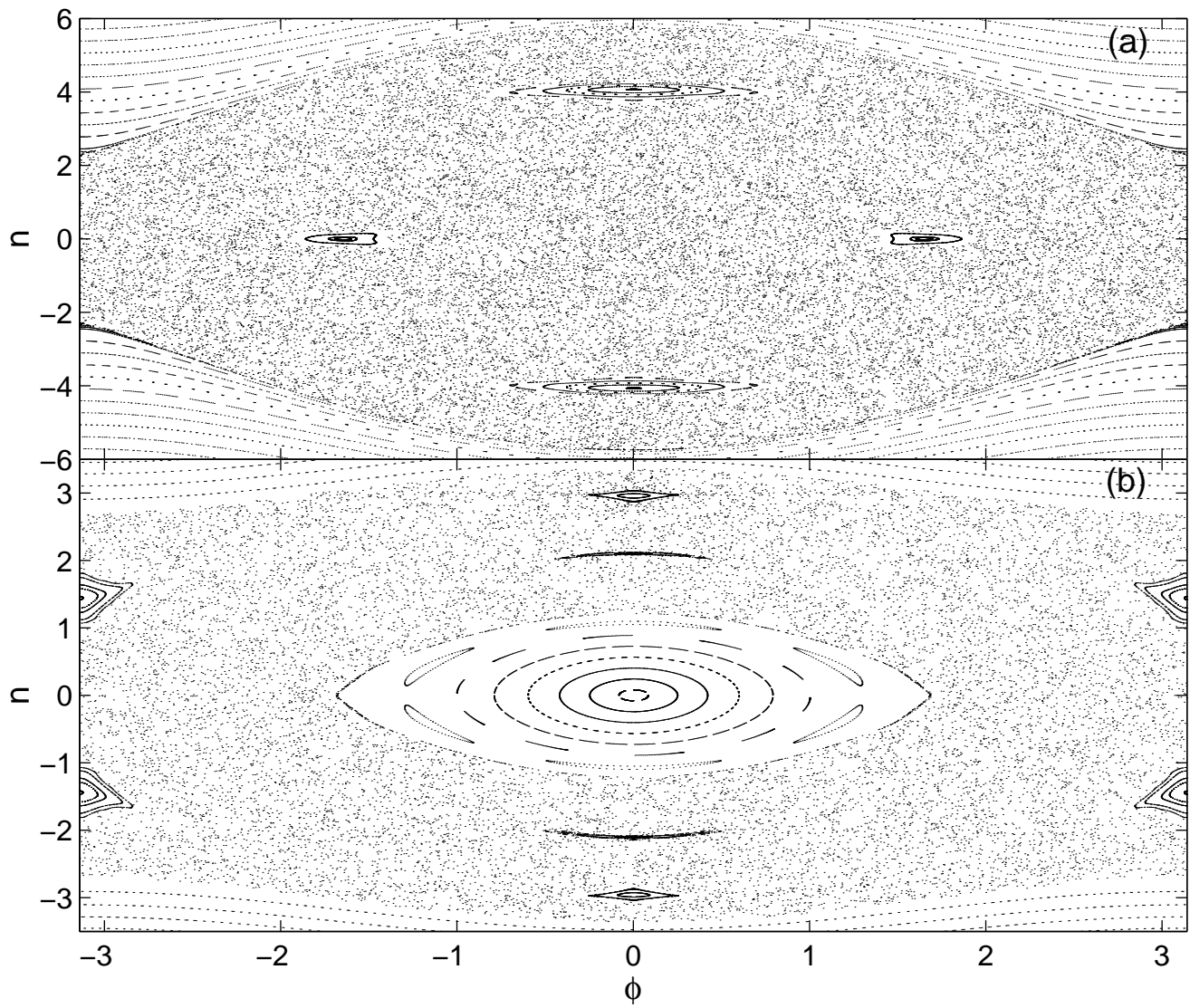
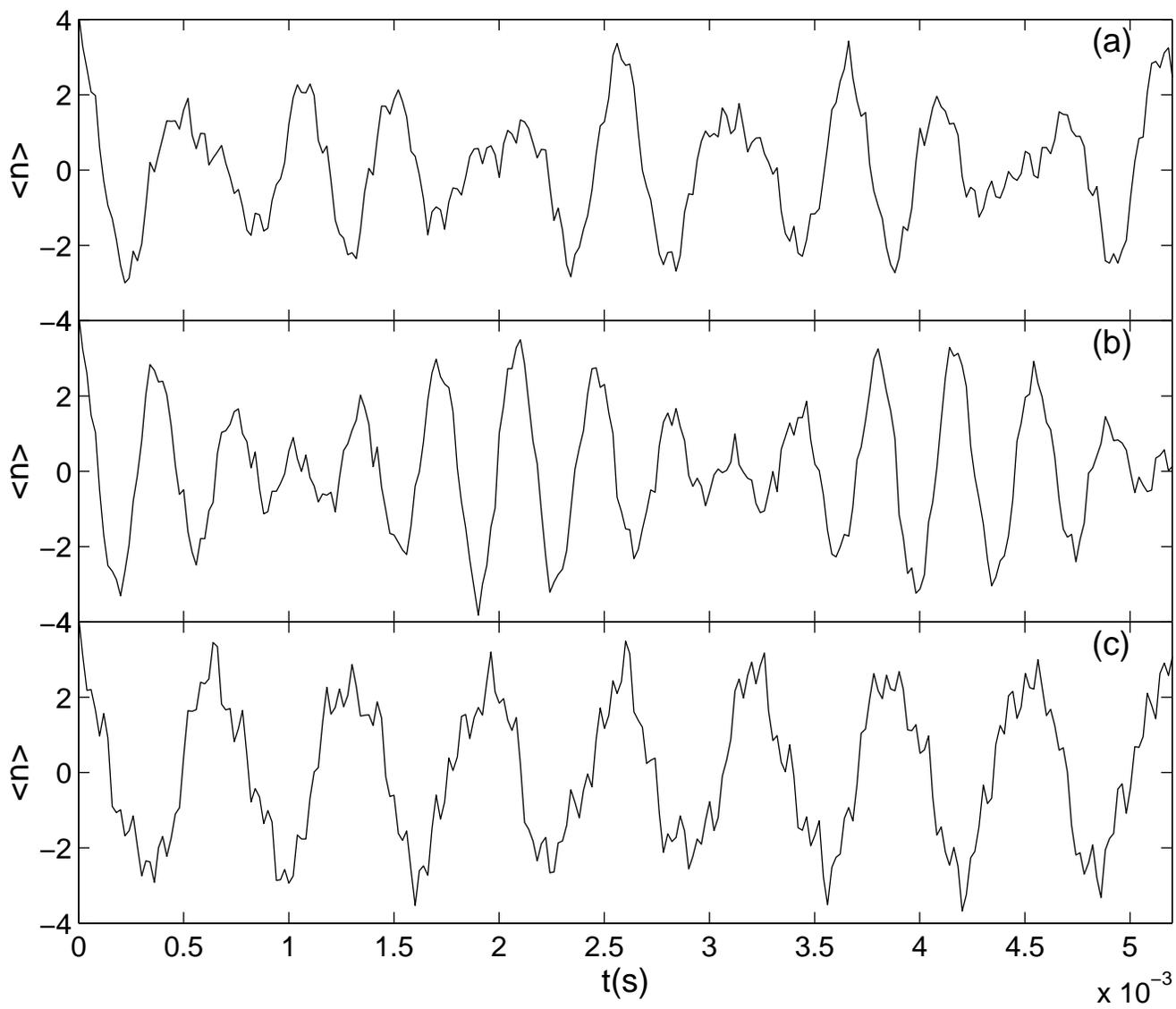
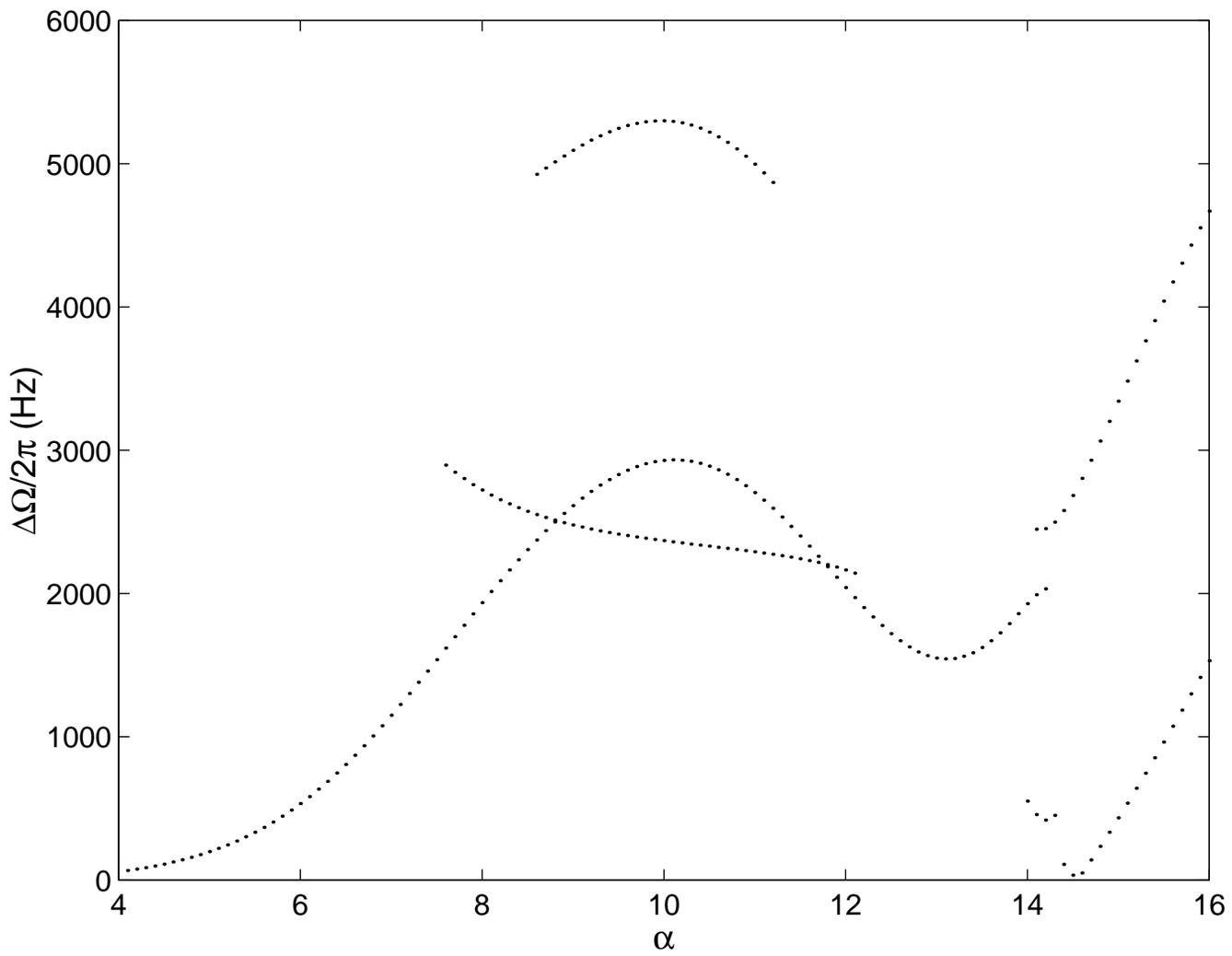
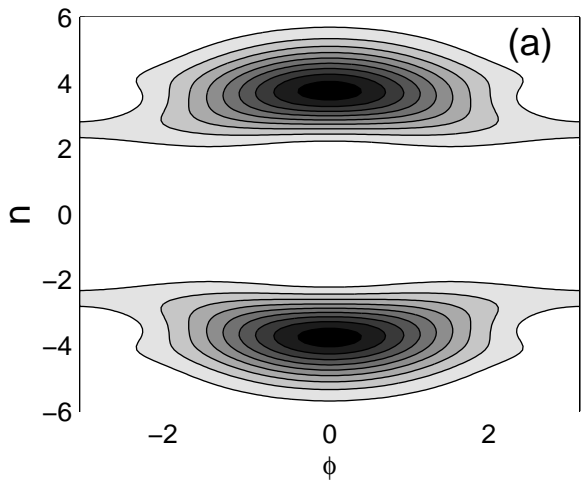
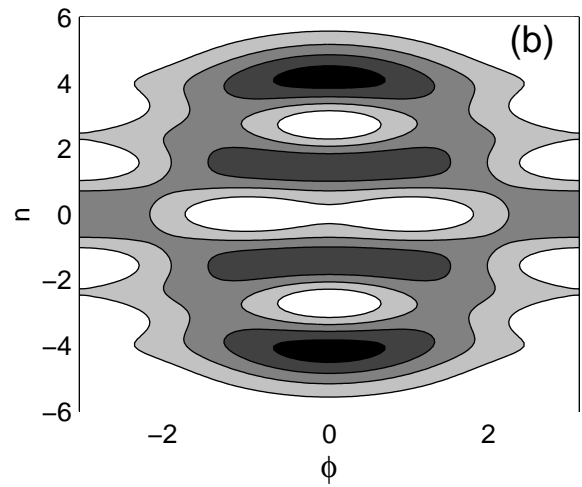
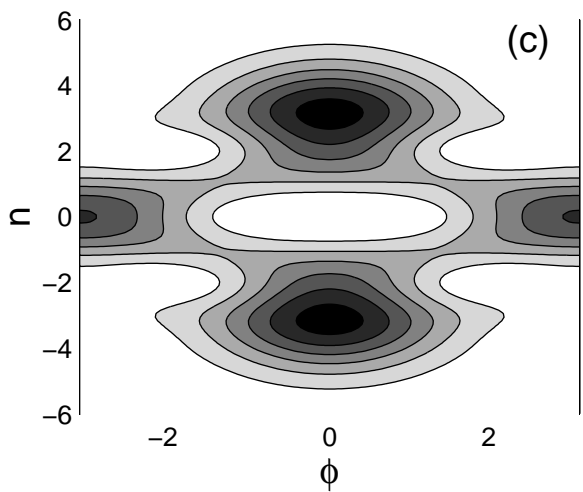
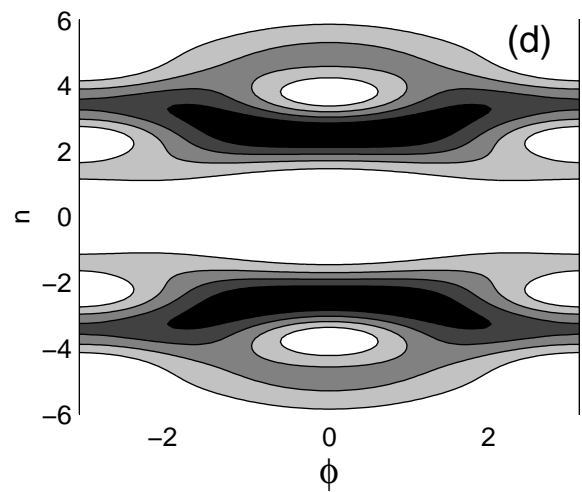
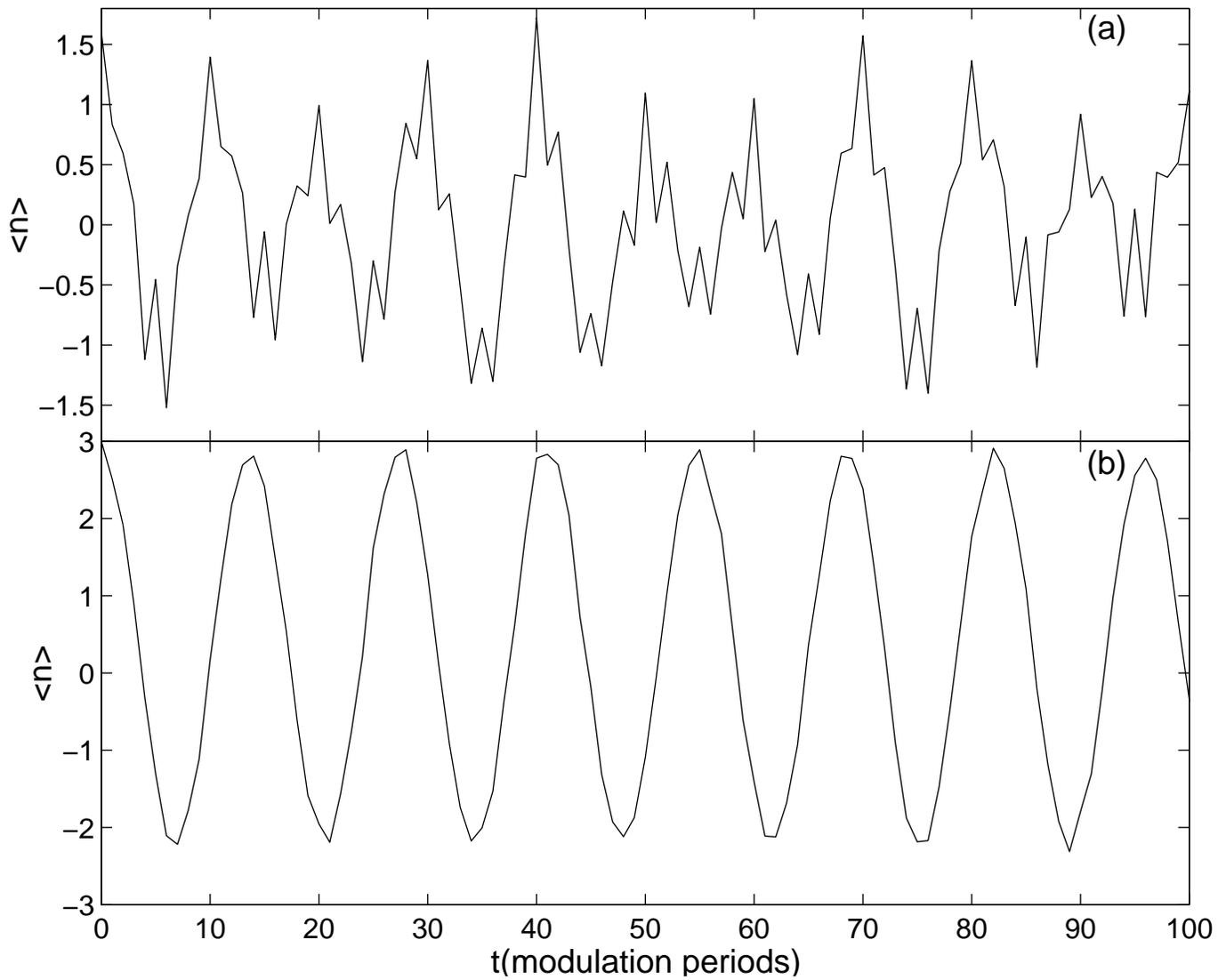
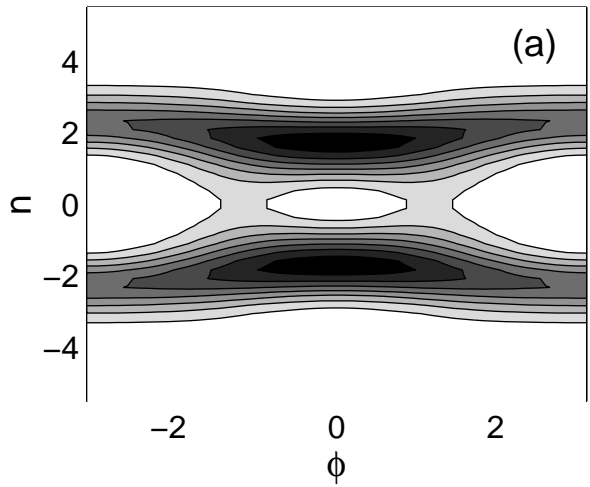
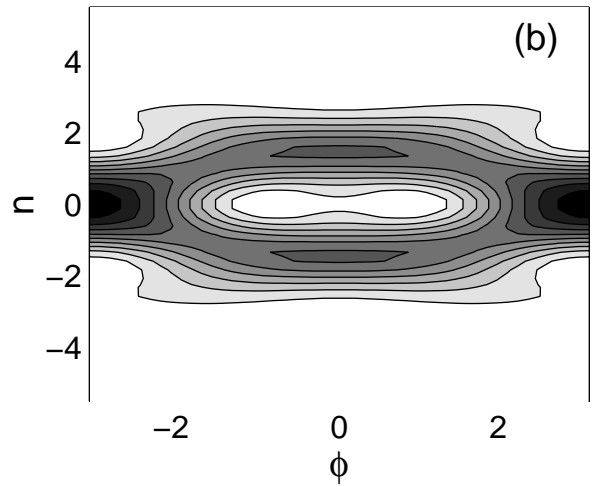
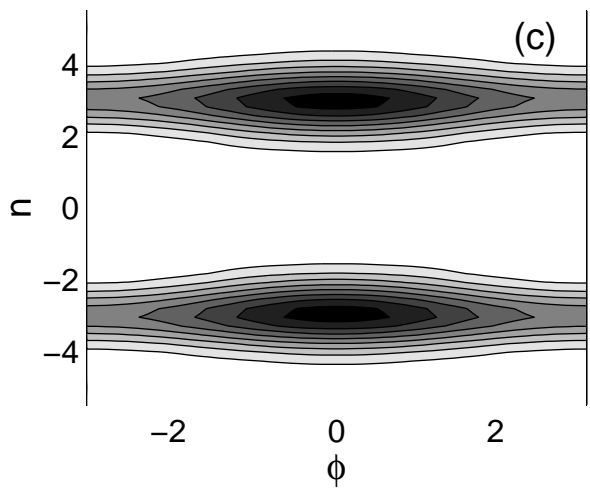
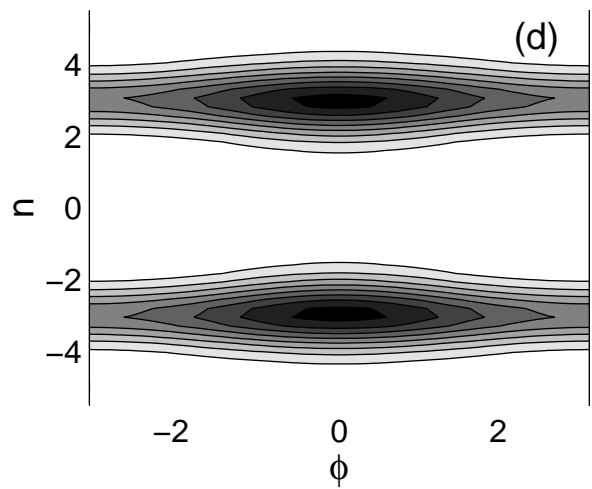
\end{document}